\documentclass[prd,aps,twocolumn,amsmath,amssymb,nofootinbib,preprintnumbers]
{revtex4}

\usepackage{graphicx}
\usepackage{dcolumn}
\usepackage{bm}
\usepackage{amsmath}
\usepackage{amsfonts}
\usepackage{euscript,bbm}
\usepackage{ifthen}
\usepackage{psfrag}
\usepackage{slashed}

\def\ls{\mathrel{\lower4pt\vbox{\lineskip=0pt\baselineskip=0pt
           \hbox{$<$}\hbox{$\sim$}}}}
\def\gs{\mathrel{\lower4pt\vbox{\lineskip=0pt\baselineskip=0pt
           \hbox{$>$}\hbox{$\sim$}}}}
\def\drawbox#1#2{\hrule height#2pt

\hbox{\vrule width#2pt height#1pt \kern#1pt
              \vrule width#2pt}
              \hrule height#2pt}

\def\Asym#1#2{\vcenter{\vbox{\drawbox{#1}{#2}
              \kern-#2pt       
              \drawbox{#1}{#2}}}}

\newcommand{\be}{\begin{equation}}
\newcommand{\ee}{\end{equation}}
\newcommand{\bea}{\begin{eqnarray}}
\newcommand{\eea}{\end{eqnarray}}

\begin{document}

%
\title{Baryogenesis and Late-Decaying Moduli}

\author{Rouzbeh Allahverdi$^{1}$}
\author{Bhaskar Dutta$^{2}$}
\author{Kuver Sinha$^{2}$}

\affiliation{$^{1}$~Department of Physics and Astronomy, University of New Mexico, Albuquerque, NM 87131, USA \\
$^{2}$~Department of Physics, Texas A\&M University, College Station, TX 77843-4242, USA}


\begin{abstract}

Late-decaying string moduli dilute the baryon asymmetry of the universe created in any previous era.
The reheat temperature for such moduli is below a GeV, thus motivating baryogenesis at very low temperatures.
We present an extension of the minimal supersymmetric standard model with TeV-scale colored fields that can yield the correct baryon asymmetry of the universe in this context.
Modulus decay, which reheats the universe at a temperature below ${\rm GeV}$, produces the visible sector fields and neutralino dark matter in non-thermal fashion. We discuss various possibilities for baryogenesis from TeV scale colored fields and show that they can generate an acceptable baryon asymmetry, while being compatible with phenomenological constraints like neutron-antineutron oscillation.

\end{abstract}
MIFPA-10-19\\ May, 2010
\maketitle


\section{Introduction}

Compactification of string theory to four dimensions produces a large number of moduli fields corresponding to complex structure and Kahler deformations. Moduli stabilization has been a major area of research in string phenomenology \cite{Douglas:2006es}. In the early universe the moduli are typically displaced form the minimum of their potential. They start oscillating about the minimum and behave like non-relativistic matter once the Hubble expansion rate drops below their mass. The moduli are long lived because their couplings to other fields are gravitationally suppressed, and hence can dominate the energy density of the universe. Their late decay can spoil the successful predictions of Big-Bang Nucleosynthesis (BBN). This is the so-called cosmological moduli problem.

Moduli with masses above $20$ TeV do not pose this problem, as they decay before primordial nucleosynthesis. 
They result in a very low reheat temperature, which is below a GeV. 
While avoiding trouble for BBN, they may still leave their imprint on cosmology, for example by non-thermal production of dark matter \cite{KMY,Watson:2009hw,Dutta:2009uf,Acharya:2009zt}.

An important issue in scenarios with late-decaying moduli is obtaining the correct baryon asymmetry of the universe. The decay of the modulus generates a large amount of entropy, which considerably dilutes any asymmetry that was created in a previous era. Affleck-Dine mechanism \cite{Affleck:1984fy} can in principle produce an $ \mathcal{O}(1)$ baryon asymmetry and yield the desired value after late-time dilution by modulus decay. A crucial element in concrete realizations of Affleck-Dine baryogenesis in effective $D=4, \, \, N=1$ supergravity is to avoid a positive Hubble-induced correction to the flat direction mass term~\cite{Dine:1995kz}. This generally requires specific couplings in the Kahler potential between the inflationary and visible sectors.

One possibility is to produce the baryon asymmetry after modulus decay. Generating sufficient asymmetry at such low temperatures is a challenging task since sphaleron transitions are exponentially suppressed, thus rendering motivated scenarios like electroweak baryogenesis and leptogenesis inapplicable.

One can consider the production of baryon asymmetry by the direct decay of moduli, through $CP$ and baryon number violating couplings to baryons. However, $R$-parity conservation can place severe constraints on such couplings. For example, consider a scenario with the superpotential couplings ${\lambda}T u^c d^c d^c/M_{\rm P}$ \cite{Thomas:1995ze,Kitano:2008tk} (also see~\cite{Cline}), where $T$ is a modulus and $u,~d$ are the right-handed up- and down-type quarks respectively ($M_{\rm P} = 2.4 \times 10^{18}$ GeV is the reduced Planck mass). Since $u^c d^c d^c$ is odd under $R$-parity, the Lightest Supersymmetric Particle (LSP) will be unstable unless $\langle T \rangle = 0$ at the minimum. However, many moduli arising from string theory are in fact stabilized at a non-zero Vacuum Expectation Value (VEV) with Planckian size. One may relax $R$-parity conservation, in which case new dark matter candidates will be needed other than the LSP.


In this paper, we will address the issue of baryogenesis in the presence of late-decaying moduli, leaving the modulus and inflationary sectors unconstrained.
The visible sector consists of the Minimal Supersymmetric Standard Model (MSSM) fields along with some additional fields that have baryon number violating couplings to MSSM fields. The $C$ and $CP$ violating decay of these additional fields, together with the out-of-equilibrium modulus decay, satisfy all the Sakharov conditions~\cite{Sakharov}. $R$-parity conservation leads to a stable LSP as the dark matter candidate. The entire scenario can take place at a reheat temperature of $10~ {\rm MeV} - 1~ {\rm GeV}$, produced by the decay of a heavy modulus. In fact, this scenario can work for a general model with low reheat temperature irrespective of the origin of the late decay.

For the visible sector, we will primarily consider an MSSM extension with $N_X$ flavors of iso-singlet color triplets $X,~{\bar X}$ and singlets $N$.
For $N_X > 1$, the desired baryon asymmetry can be obtained with the help of the self-energy correction in $X,~{\bar X}$ decays. For $N_X = 1$, we require multiple singlets whose decays can result in the baryon asymmetry via vertex and self-energy corrections. One can also consider MSSM extensions without singlets, for example, with fields $Y$ that transform as $(3,2)$ under $SU(3)_C$ and $SU(2)_W$.
Our discussions do not depend crucially on the details of the modulus sector. For illustration we will take KKLT compactification in type IIB string theory as a concrete example.

The paper is organized as follows. In Section 2 we review the cosmological history for a generic scenario with late-decaying modulus. In Section 3 we describe the visible sector of the model and baryogenesis. In Section 4 we briefly discuss some issues related to dark matter and phenomenology in this model. We close the paper by concluding remarks in Section 5. An explicit example of the modulus sector along with a detailed discussion of modulus decay modes and their branching ratios are given in the Appendix.



\section{Cosmological History}


We first describe the generic cosmological history in models with late-decaying moduli. The modulus $T$, with mass $m_T$, is typically displaced from the minimum of its potential in the early universe. For example, this can happen due to quantum fluctuations during inflation~\footnote{The modulus itself could be the inflaton, as happens in examples of Kahler moduli inflation based on KKLT or racetrack inflation (for example, see~\cite{ADS}.}. The field $T$ remains frozen so long as $H > m_T$. After reheating the universe is radiation dominated, and hence $T$ starts oscillating about its minimum at a temperature $T_{\rm osc} \sim \sqrt{m_T M_{\rm P}}$. The coherent oscillations of $T$ behave like non-relativistic matter, which implies that the ratio of the energy density in the modulus $\rho_T$ to that of radiation $\rho_{\rm rad}$ increases as $R$ (with $R$ being the scale factor of the universe). The modulus will eventually dominate the energy density of the universe, and later it decays and reheats the universe \cite{Moroi}.

The decay rate of the modulus is (the decay modes have been discussed in the Appendix)
\be \label{dec}
\Gamma_{T} = \frac{c}{2\pi} \frac{m_T^3}{M^2_{\rm P}} ,
\ee
where $c \sim 0.4$. The reheat temperature of the universe after modulus decay is $T_{\rm R} \sim \left(\Gamma_T M_{\rm P}\right)^{1/2}$, which after using Eq.~(\ref{dec}) reads
\begin{eqnarray}\label{Tr}
T_{\rm R} \sim c^{1/2} \left(\frac{10.75 }{g_*}\right)^{1/4}
\left( \frac{m_T}{100\, {\rm TeV}}\right)^{3/2}\, 6.37\, {\rm MeV}.
\end{eqnarray}
where $g_*$ denotes the number of relativistic degrees of freedom. For a modulus mass $m_T \sim (1000-3000)$ TeV, we obtain $T_{\rm R} \sim (200-500)$ MeV. Thus modulus decay occurs sufficiently early not to destroy the success of BBN.

One has to be careful about other late-decaying particles, notably gravitinos, because they may also affect temperature of the universe. Modulus decay generates huge entropy that suppresses the number of gravitinos produced from reheating after inflation and renders them irrelevant. But gravitinos are also produced during the second stage of reheating by modulus decay. Thermal production of gravitino is suppressed because of the very low reheat temperature. However, gravitinos are also directly produced from the modulus decay.

Let us denote the branching ratio for the decay of modulus to a pair of gravitinos by ${\rm Br}_{\rm gravitino}$ (for details, see Appendix). The fraction of energy density in the gravitinos right after modulus decay is ${\rm Br}_{\rm gravitino}$. Gravitinos have an energy $(m_T/2) > m_{3/2}$, and hence are relativistic, upon production. The fractional energy density of gravitinos remains equal to ${\rm Br}_{\rm gravitino}$ until transition to the non-relativistic regime. This happens when temperature of the universe is
%
$\sim \left({m_{3/2}/m_T}\right) T_{\rm R}$.
%
From this moment on the energy density in gravitinos $\rho_{3/2}$ is redshifted $\propto R^{-3}$.
Using the gravitino decay rate $\Gamma_{3/2} \sim m^3_{3/2}/M^2_{\rm P}$, we can find temperature of the universe at the time of gravitino decay $T_{3/2} \sim (m^3_{3/2}/M_{\rm P})^{1/2}$. Then, since temperature is redshifted $\propto R^{-1}$, we have
\be \label{gravdec}
{\rho_{3/2} \over \rho_{\rm rad}} \sim \left({m_{3/2} \over m_T}\right) \left({T_{\rm R} \over T_{3/2}}\right) {\rm Br}_{\rm gravitino} ,
\ee
at the time of gravitino decay. For $m_{3/2} > 40$ TeV gravitinos decay before BBN (i.e., $T_{3/2} > 1$ MeV). For $m_T \sim (1000-3000)$ TeV, which implies $T_{\rm R} \sim (200-500)$ MeV, gravitinos carry a small fraction of the total energy density when they decay provided that ${\rm Br}_{\rm gravitino} \ll 1$. As worked out in the Appendix, one obtains ${\rm Br}_{\rm gravitino} \sim 0.01$ for the modulus sector we have considered. Therefore gravitino decay produces negligible entropy that does not affect temperature of the universe.

\section{Baryogenesis from modulus decay}

As described above, we will consider baryogenesis at low reheat temperatures $T_{\rm R} \simeq 10 {\rm MeV} - 1 {\rm GeV}$, resulted from the decay of a modulus that couples gravitationally to the visible sector. In this section, we will describe a particular model that can achieve this.
The origin of the modulus does not have any bearing on our results, thus we assume a generic scenario with late-decaying modulus.
An explicit example of the modulus sector is given in the Appendix.

\subsection{Baryogenesis from color triplets}

The visible sector is the same as MSSM augmented with extra fields: $N_X$ flavors of iso-singlet color triplets $X_{\alpha},~\bar{X}_{\alpha}$ ($1 \leq \alpha \leq N_X$) with hypercharges $+4/3,-4/3$ respectively, and a singlet $N$. We emphasize that $N$ is singlet under the Standard Model (SM) gauge symmetry, but may be charged under a larger gauge group. The superpotential is $W = W_{\rm MSSM} + W_{\rm extra}$, where
\bea\label{superpot1}
W_{\rm extra} &=& \lambda_{i \alpha} N u^c_i X_{\alpha} + \lambda^\prime_{ij\alpha} d^c_i d^c_j
\overline{X}_{\alpha} \\ \nonumber
&+& {M_N \over 2} NN + M_{\alpha} X_{\alpha} \overline{X}_{\alpha}~.
\eea
Here $i,~j$ denote MSSM flavor indices (color indices are suppressed for simplicity). We note that $\lambda^\prime_{ij\alpha}$ is antisymmetric under $i \leftrightarrow j$. We assign charges $+1$ and $-1$, respectively, to the fermionic components of $N$ and $X,~{\bar X}$ under $R$-parity. This insures $R$-parity conservation, and hence LSP as the dark matter candidate.

The mass eigenvalues for the scalar components of a given flavor are given by
\be \label{mass}
m^2_\alpha = |M_{\alpha}|^2 + \tilde{m}_{\alpha}^2 \pm \vert B_\alpha M_{\alpha} \vert ,
\ee
where ${\tilde m}_\alpha$ is the soft mass of scalars and $B_\alpha$ is the $B$-term associated with the superpotential mass term $M_\alpha X_\alpha {\bar X}_\alpha$.

We now consider generation of the Baryon Asymmetry of the Universe (BAU) from the decay of color triplets~\footnote{A different model for baryogenesis from color triplets, with broken $R$-parity, has been studied in~\cite{Cohen:2009tx}.}. The first step is to find the yield of $X_{\alpha},~\overline{X}_{\alpha}$ from modulus decay. As outlined in the Appendix, the modulus mainly decays to gauge bosons and gauginos (branching ratio $\sim 0.98$), as well as $X_{\alpha},~\overline{X}_{\alpha}$ (branching ratio $\sim 0.01$) and gravitinos (branching ratio $\sim 0.01$). Since $X_\alpha,~{\bar X}_\alpha$ are colored, their scalar and fermionic components will also be produced from the decay of gluinos. The branching ratio for this mode is $\sim 0.1$ (if the squarks are lighter than glunios), and can be ${\cal O}(1)$ if (some of) the squarks are heavier than gluinos.


We denote the branching ratio for production of scalar and fermionic components of $X_\alpha,~{\bar X}_\alpha$ from $T$ decay by ${\rm Br}_\alpha$.
Then, after using Eq.~(\ref{Tr}), we find:
\begin{eqnarray}
Y_{\alpha} & = & Y_T ~ {\rm Br}_\alpha = \frac{3}{4} \frac{T_{\rm R}}{m_T} ~ {\rm Br}_\alpha\; , \nonumber \\
& \sim & 4 \times 10^{-8} \left(\frac{m_T}{100 \rm{TeV}}\right)^{\frac{1}{2}} ~ {\rm Br}_\alpha \, ,
\end{eqnarray}
where $Y_T = 3T_{\rm R}/4m_T$ is the dilution factor for $T$ decay, and $Y_\alpha$ denotes the ratio of (the common value of) the number density of $X_\alpha,~\psi_\alpha,~{\bar X}_\alpha,~{\bar \psi}_\alpha$ (and their antiparticles) to the entropy density $s$. Here $\psi,~{\bar \psi}$ denote the fermionic components of $X,~{\bar X}$ superfields (we use the same symbol for superfields and their scalar components). Considering a typical modulus mass $m_T \sim 1000$ TeV, we have $Y_\alpha \sim 1.25 \times 10^{-7} ~ {\rm Br}_\alpha$.

The BAU is then given by
\be \label{BAU}
\eta_B \equiv \frac{n_B - n_{\overline{B}}}{n_{\gamma}} = 7.04 ~ \sum_{\alpha} Y_\alpha {\epsilon_{\alpha}},
\ee
where $\epsilon_\alpha$ is the generated asymmetry per flavor of $X,~{\bar X}$.

To demonstrate baryogenesis from the decay of $X,~{\bar X}$, we specifically consider two flavors of color triplets ($N_X = 2$), which is the minimum number required to obtain an asymmetry. We focus on decays governed by supersymmetry conserving interactions. Let us first consider the decay of fermionic components, starting with $\psi_1$. The relevant decay modes (if kinematically open) are ${\bar \psi}_1 \rightarrow d^{c*}_i {\tilde d}^{c*}_j$, for which $\Delta B = +2/3$, and ${\bar \psi}_1 \rightarrow {\tilde N} u^c_k,~N {\tilde u}^c_i$ with $\Delta B = -1/3$. Here $N,{\tilde N}$ denote the fermionic and scalar components of $N$ superfield respectively. The interference of tree-level and one-loop self-energy diagrams in Figs.~1,2 will result in a baryon asymmetry from ${\bar \psi}_1$ and ${\bar \psi}^*_1$ decays (provided that $M_1 > M_N$):
\be \label{asymmetry}
\epsilon_1 = {1 \over 8 \pi} ~ {\sum_{i,j,k} {\rm Im} \left(\lambda^*_{k1}\lambda_{k2}\lambda^{\prime *}_{ij1}\lambda^{\prime}_{ij2}\right) \over \sum_{i,j}\lambda^{\prime *}_{ij1} \lambda^{\prime}_{ij1} + \sum_{k}\lambda^*_{k1} \lambda_{k1}} ~ {\cal F}_S \left(M^2_2 \over M^2_1 \right), 
\ee
where, for $M_2 - M_1 > \Gamma_{{\bar \psi}_1}$, we have
\be \label{self}
{\cal F}_S(x) = {2 \sqrt{x} \over x - 1}.
\ee

The expression in Eq.~(\ref{asymmetry}) is similar to that in the standard leptogenesis~\cite{leptogenesis}. Note however that there are no vertex diagrams in this case.

We find the same asymmetry from ${\psi}_1$ and $\psi^*_1$ decays. This can be understood from the fact that ${\bar \psi}_1$ and $\psi^c_1$ form a four-component fermion with hypercharge quantum number $-4/3$. Also, in the limit of unbroken supersymmetry, we get exactly the same asymmetry from the decay of scalars $X_1,~{\bar X}_1$ and their antiparticles $X^*_1,~{\bar X}^*_1$. In the presence of supersymmetry breaking the asymmetries from fermion and scalar decays will be similar provided that $m_{1,2} \sim M_{1,2}$, see Eq.~(\ref{mass}).

Similarly, for $M_2 > M_N$ the decay of the scalar and fermionic components of $X_2,{\bar X}_2$ will result in an asymmetry $\epsilon_2$, which follows the same expression as in~(\ref{asymmetry}) with $1 \leftrightarrow 2$. We therefore find:
\bea\label{etaB}
\eta_B = 7.04 \times 10^{-6} ~ {1 \over 8 \pi} ~ {M_1 M_2 \over {M^2_2 - M^2_1}} ~ \sum_{i,j,k} {\rm Im} \left(\lambda^*_{k1}\lambda_{k2}\lambda^{\prime *}_{ij1}\lambda^{\prime}_{ij2}\right) \, \nonumber \\
\left[{{\rm Br}_1 \over \sum_{i,j}\lambda^{\prime *}_{ij1} \lambda^{\prime}_{ij1} + \sum_{k}\lambda^*_{k1} \lambda_{k1}} + {{\rm Br}_2 \over \sum_{i,j}\lambda^{\prime *}_{ij2} \lambda^{\prime}_{ij2} + \sum_{k}\lambda^*_{k2} \lambda_{k2}} \right] \, . \nonumber \\
\,
\eea
%
For ${\rm Br}_1 \sim {\rm Br}_2 \sim 0.1$\ and $M_1 \sim M_2$, we need
\be
\sum_{i,j,k} {\rm Im} \left(\lambda^*_{k1}\lambda_{k2}\lambda^{\prime *}_{ij1}\lambda^{\prime}_{ij2}\right) \left({1 \over \sum_{i} \lambda^*_{i1} \lambda_{i1}} + {1 \over \sum_{i} \lambda^*_{i2} \lambda_{i2}}\right) \sim 0.01 ,
\ee
in order to obtain the canonical value $4 \times 10^{-10} \leq \eta_B \leq 7 \times 10^{-10}$. Assuming
similar couplings to all flavors of (s)quarks such that $|\lambda_{i1}| \sim |\lambda_{i2}| \gg |\lambda^{\prime}_{ij1}| \sim |\lambda^{\prime}_{ij2}|$ ($1 \leq i,j \leq 3$), and for $CP$ violating phases of ${\cal O}(1)$ in $\lambda$ and $\lambda^{\prime}$, this can be achieved for
\be \label{lambda}
|\lambda_{i1}| \sim |\lambda_{i2}| \sim 1 ~ ~ , ~ ~ |\lambda^{\prime}_{ij1}| \sim |\lambda^{\prime}_{ij2}| \sim 0.04 .
\ee
For $|\lambda_{i1}| \sim |\lambda_{i2}| \sim |\lambda^{\prime}_{ij1}| \sim |\lambda^{\prime}_{ij2}|$, we need couplings $\sim 0.1$ to generate the correct asymmetry. One can also obtain the desired $\eta_B$ for smaller couplings with the help of a mild degeneracy between the triplets $M_2 - M_1 \ls M_1$.

\begin{figure}[ht]
\centering
\includegraphics[width=3.5in]{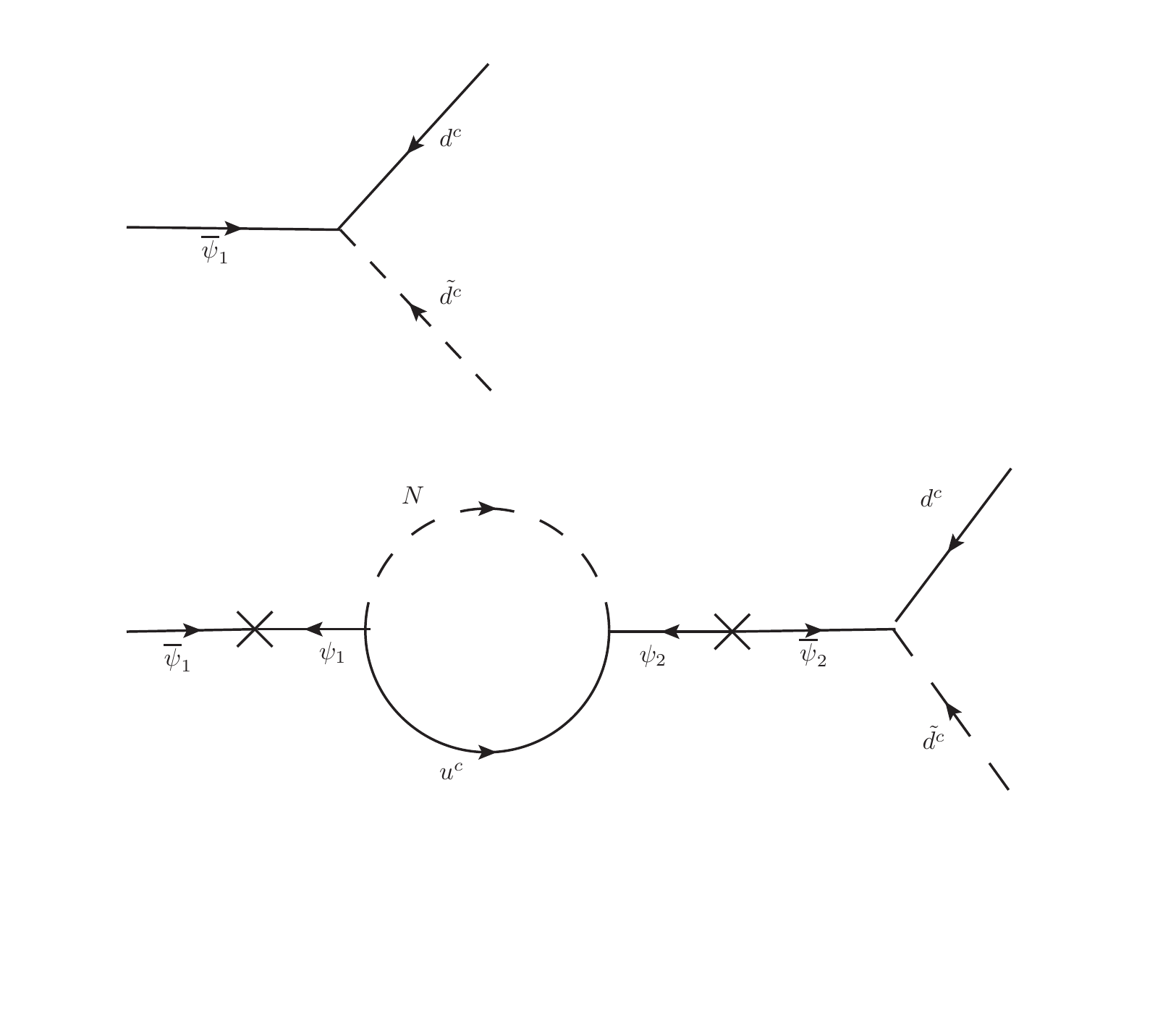}
\caption{Tree-level and self-energy diagrams for the decay ${\bar \psi}_1 \rightarrow d^{c*} {\tilde d}^{c*}$.}
\end{figure}

\begin{figure}[ht]
\centering
\includegraphics[width=3.5in]{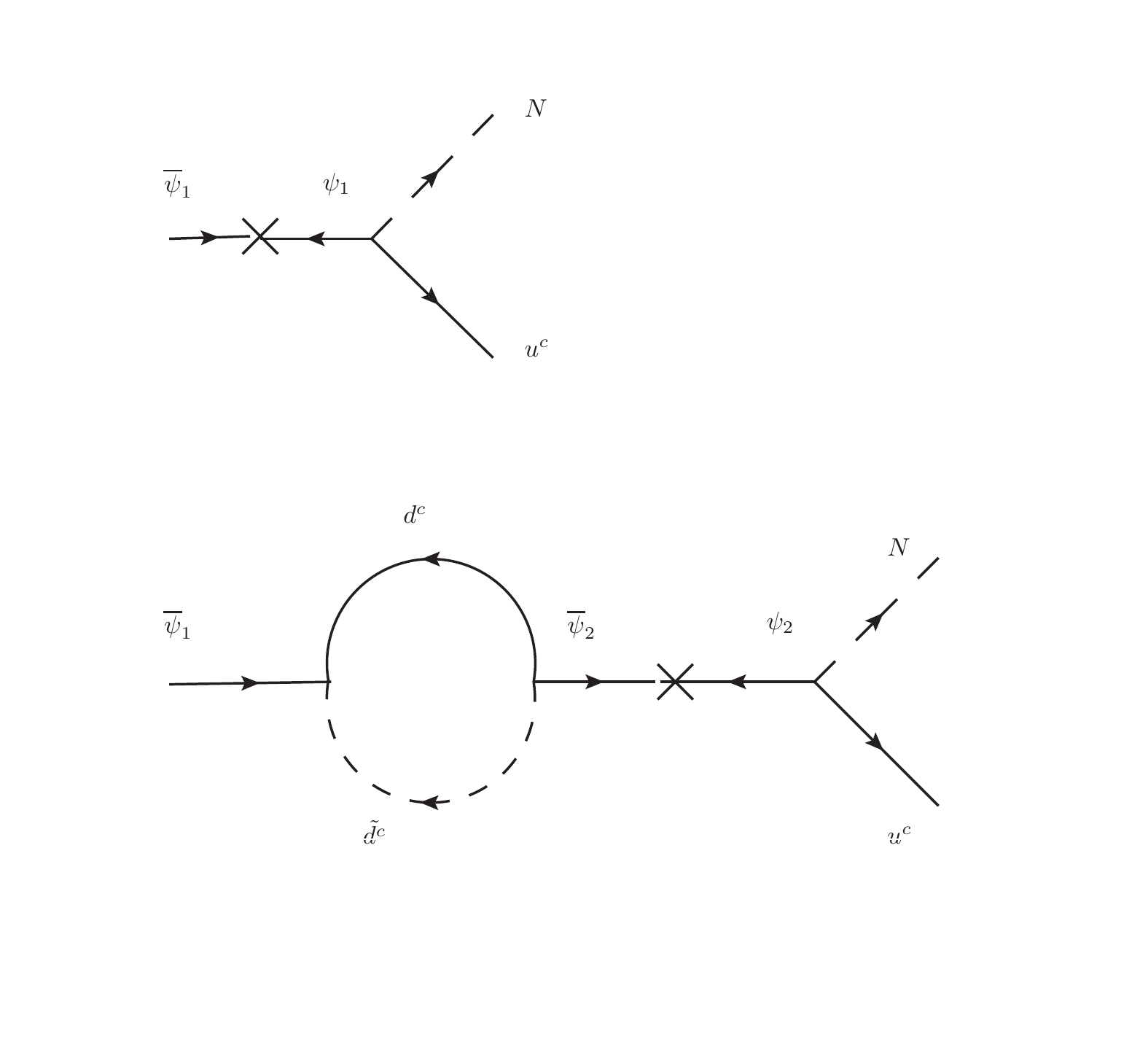}
\caption{Tree-level and self-energy diagrams for the decay ${\bar \psi}_1 \rightarrow N u^c$.}
\end{figure}

We note that there are contributions to the baryon asymmetry coming from decay diagrams that include supersymmetry breaking interactions (similar to the soft leptogenesis scenario~\cite{soft}). We do not consider these diagrams here.

\subsection{Baryogenesis from singlets}

In the case where there is a single flavor of color triplets $X, \overline{X}$, decay diagrams in Figs.~1,2 will not give rise to any baryon asymmetry. We then need to consider variations of the superpotential in Eq.~(\ref{superpot1}), for example, by introducing multiple singlets $N_{\alpha}$:
\bea \label{superpot2}
W_{\rm extra} &=& \lambda_{i\alpha} N_{\alpha} u^c_i X + \lambda^\prime_{ij} d^c_id^c_j \overline{X} \\ \nonumber
&+& {M_{\alpha} \over 2} N_{\alpha} N_{\alpha} + M_{X} X \overline{X} \, .
\eea
In this case the interference between tree-level and one-loop self-energy and vertex diagrams, like those shown in Fig.~3, will result in a baryon asymmetry (provided that $M_\alpha > M_X$). For example, the asymmetry generated in $N_\alpha \rightarrow X^* u^{c*}_i$ decay (with $N_\alpha$ being the fermionic component of the corresponding superfield) is given by:

\begin{figure}[ht]
\centering
\includegraphics[width=3.5in]{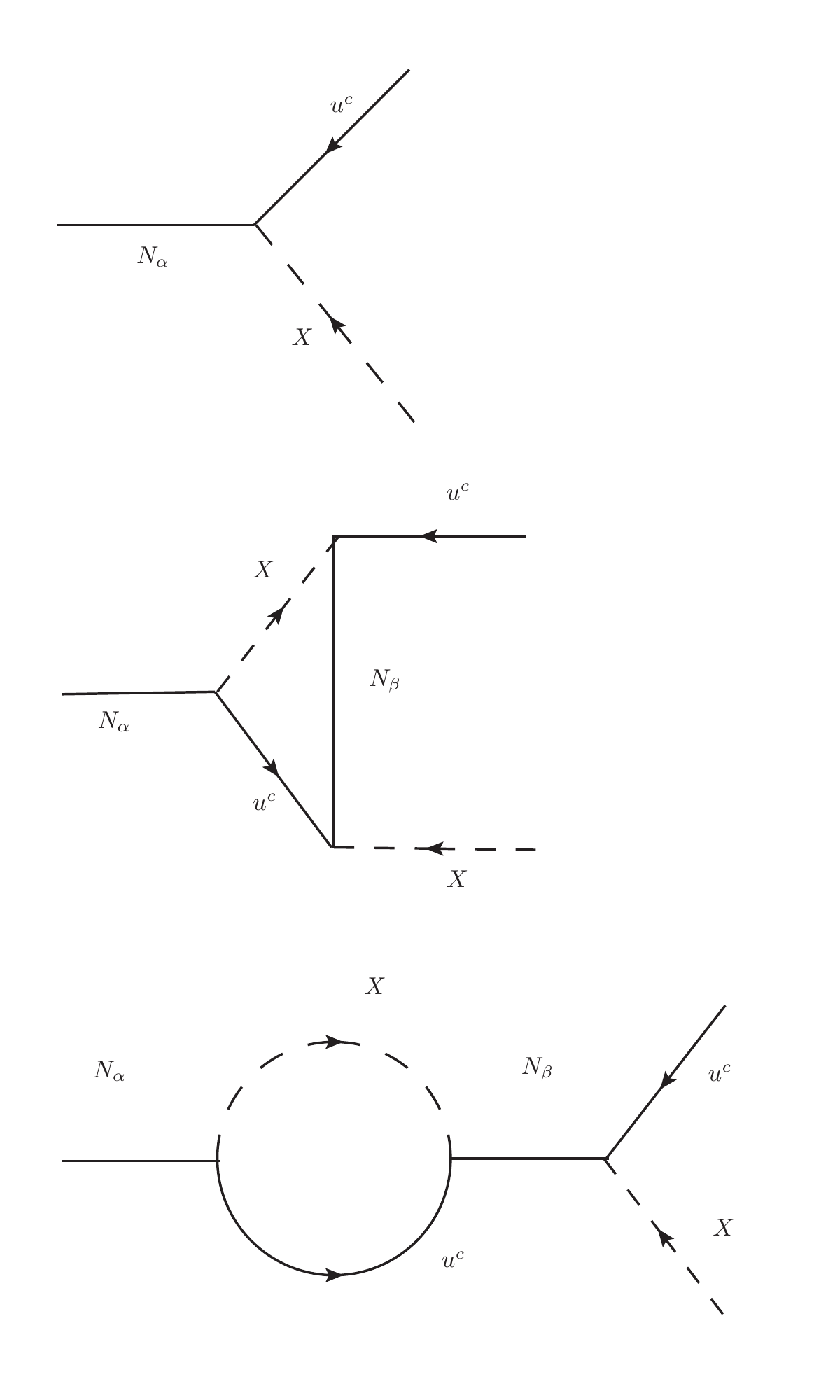}
\caption{Tree-level, self-energy and vertex diagrams for the decay $N_\alpha \rightarrow X^* u^{c*}$.}
\end{figure}

%
\be \label{Nasymmetry}
\epsilon_\alpha = {\sum_{i,j,\beta} {\rm Im} \left(\lambda_{i \alpha} \lambda^*_{i \beta}\lambda^{*}_{j \beta}\lambda_{j\alpha}\right) \over 24 \pi \sum_{i}\lambda^{*}_{i\alpha} \lambda_{i\alpha}} ~ \left[3{\cal F}_S \left(M^2_\beta \over M^2_\alpha \right) + {\cal F}_V \left(M^2_\beta \over M^2_\alpha \right)\right] ,
\ee
where
\be
{\cal F}_S (x) = {2 \sqrt{x} \over x - 1} ~ ~ , ~ ~ {\cal F}_V = \sqrt{x} ~ {\rm ln} \left(1 + {1 \over x}\right) .
\ee
This is the same expression as that for asymmetry in the standard leptogenesis~\cite{leptogenesis}, with an additional factor of $3$ in the denominator because the final state has baryon number $+1/3$. The factor 3 in the self energy contribution arises due to the sum over colors in intermediate states. The total asymmetry is found by summing up over contributions from fermionic and scalar components of all flavors of $N$. Again, one can obtain the desired BAU by suitable choices of $\lambda_{i \alpha}$.

One comment is in order. Since $N_\alpha$ are singlet under the SM gauge group, they are not produced from the decay of MSSM fields (notably gluinos) unlike $X,~{\bar X}$. However, as shown in the Appendix, they can be produced directly from modulus decay with a typical branching ratio of ${\cal }(0.01)$. Their yield $Y_N$ will therefore be generically smaller than the yield of $X,~{\bar X}$, which requires a larger asymmetry factor~(\ref{Nasymmetry}) in order to obtain the desired BAU.

\subsection{Other possibilities}

One can also have baryogenesis for other variations of the model. For example, consider the case with a single flavor of $X$ and a single flavor of $N$. In this case, the asymmetry can be generated in the three-body decay of $N$ (via off-shell $X$), with the one-loop correction arising from $W^\pm$ exchange~\cite{Babu:2006wz}:
\be
\epsilon \sim \left({-\alpha_2 \over 4}\right) \left({m_c m_t m_s m_b \over m_W^2 M_N^2} \right),
\ee
where $m_W$ is the $W$ mass ansd $\alpha_2$ is the $SU(2)_W$ fine structure constant. Taking $M_N \sim 10$ GeV, so that $N$ decay can produce bottom quarks, we find $\epsilon \sim 10^{-6}$.
We therefore need $Y_N \sim 10^{-4}$ in order to obtain the correct BAU. This requires a very high modulus mass $m_{T} \sim 10^{8}$ TeV, for which the reheat temperature~(\ref{Tr}) will be above the electroweak scale.

Another possibility is to rely on supersymmetry breaking interactions, similar to what happens in soft leptogenesis~\cite{soft}. The baryon asymmetry can then be generated in the decay of $X,~{\bar X}$ or ${\tilde N}$ decay. As in the soft leptogenesis scenario, creation of the asymmetry does not need multiple flavors of $X,~{\bar X}$ or $N$.

Finally, we can have baryogenesis without any singlets.
In this case additional fields with MSSM gauge charges will replace $N$. For example, consider iso-doublet color triplet fields $Y,~\overline{Y}$ that have hypercharge numbers $\mp 5/3$ respectively. The relevant superpotential terms then read
\bea
W_{\rm extra} &=& \lambda_{i\alpha} Y Q_i X_{\alpha} + \lambda^\prime_{ij\alpha} d^c_id^c_j
\overline{X}_{\alpha} \\ \nonumber
&+& M_Y Y \overline{Y} + M_{\alpha} X_{\alpha} \overline{X}_{\alpha}~.
\eea
The BAU will be generated in the decay of $X_\alpha,~{\bar X}_\alpha$ through diagrams similar to those in Figs.~1,2 but $N,~u^c_i$ replaced with $Y,~Q_i$.

\section{Comments on Dark Matter and Phenomenology}

Since there are no $R$-parity violating terms in the superpotential, the LSP is absolutely stable. The superparticles in our model include the new fermions $\psi_\alpha,~{\bar \psi}_{\alpha}$ of the color triplet superfields $X, \overline{X}$ and the scalar(s) $N$, along with the usual MSSM superpartners.

For the particular modulus sector chosen, see the Appendix, the soft masses arise from a combination of modulus and conformal anomaly contributions. The mirage scale is $\mu_{\rm Mir} \sim 3 \times 10^9$ GeV. The gaugino masses at a scale $\mu$ are then given by $M(\mu) = \frac{g^2(\mu)}{g^2(\mu_{\rm Mir})}M_0$. Typically, the LSP will be a Bino/Higgsino in this case.

Dark matter is produced non-thermally, since the reheat temperature is very low, from modulus decay. Its annihilation cross section must be enhanced relative to the nominal value in thermal scenarios $\langle \sigma_{\rm ann} v \rangle = 3 \times 10^{-26}$ cm$^2$ in order to yield the correct relic abundance upon non-thermal production. For a LSP mass of 200 GeV, the enhancement factor is given by $(T_{\rm f}/T_{\rm R}) \sim 50$, where $T_{\rm f} \sim 10$ GeV is the freeze-out temperature, and $T_{\rm R} \sim 200$ MeV is the reheat temperature.

The color triplets ($X$,$\bar X$) can be pair produced at the Large Hadron Collider (LHC) and the final states of these production process will contain multi jets plus multi leptons and missing energy. The jets and leptons will be produced from the cascade decays of these particles into the LSP neutralino via squarks, heavier neutralinos, charginos and sleptons. The mass scale of these triplets can be measured by measuring the effective mass $M_{\rm eff}$ of four highest $E_T$ jets and missing energy.

Because of baryon number violation, see Eq.~(\ref{superpot1}), it is possible to generate neutron-antineutron oscillations in this model.
The dimension 9 operator $G=\lambda^2_1{\lambda^2}_{12}^{\prime}(u^c d^c s^c)^2/(M^4_X M_N)$ is responsible for oscillations, which proceeds through the strange quark content of the neutron (because $\lambda^{\prime}_{ij}$ is antisymmetric under $ i \leftrightarrow j$). The oscillation time $t$ is given as $1/(2.5 10^{-5}\, G)$ sec, where $2.5\times 10^{-5}$ is the value for the hadronic form factor~\cite{shrock}. The current bound on the oscillation time $t < 0.86 \times 10^8$ sec~\cite{expt} requires that $G < 3\times 10^{-28}$ ${\rm GeV}^{-5}$. Using this bound, for $M_X \sim M_N \sim 1$ TeV, we find $(\lambda_1 ~ \lambda^{\prime}_{12}) < 10^{-4}$. If we use flavor universal values for $\lambda_i$ and $\lambda^{\prime}_{ij}$, then the correct amount of BAU would need a degeneracy between $M_1$ and $M_2$ at $1\%$ level. However, since all possible flavor combinations of $\lambda_k \lambda^{\prime}_{ij}$ appears in the expression for baryon asymmetry, see Eq.(\ref{etaB}), it is easy to satisfy the oscillation bound without requiring any degeneracy between $M_1$ and $M_2$ by choosing flavor nonuniversal values for $\lambda_i$ and $\lambda^{\prime}_{ij}$.


\section{Conclusion}

In this paper, we have considered baryogenesis at very low reheat temperatures. This is the generic situation in the presence of late-decaying string moduli. Heavy moduli decay before BBN and reheat the universe to a temperature below GeV. The decay releases a huge amount of entropy and dilutes any previously generated baryon asymmetry by a large factor.

One possibility is to generate BAU at this epoch, but one cannot invoke scenarios like electroweak baryogenesis and leptogenesis that rely on sphaleron processes. The baryon asymmetry may be directly produced in modulus decay to the MSSM fields. This, however, requires $R$-parity violation in which case LSP cannot be the dark matter candidate.

We instead considered baryogenesis in the visible sector. The model includes additional TeV scale fields in the visible sector: iso-singlet color triplets $X,~\overline{X}$ and either SM singlets $N$ or iso-doublet color triplets $Y,~\overline{Y}$. They are produced in modulus decay (directly and indirectly) and in turn decay to MSSM fields through baryon number violating (but $R$-parity conserving) interactions. As we saw, there are various possibilities for baryogenesis from the decay of single or multiple flavors of the color triplet and singlet fields.
For a typical modulus mass $\sim 1000$ TeV, corresponding to a reheat temperature of 200 MeV, one can obtain the correct baryon asymmetry for moderate values of couplings between the MSSM fields and extra fields. Our scenario works for a general model with low reheat temperature.

The current constraint on the neutron-antineutron oscillation time $\sim 10^8$ sec puts constraints on the couplings of color triplets to (s)quarks. For flavor nonuniversal values of these couplings one can comfortably satisfy the bound on the oscillation time and generate the correct baryon asymmetry. For flavor universal values a degeneracy at the level of $1\%$ between the color triplets is needed.

At the LHC, the color triplet fields will be pair produced and the final states of such production process will contain multi jets plus multi leptons and missing energy. The dark matter candidate in this model is a Bino/Higgsino, which is produced from modulus decay and is stable due to $R$-parity conservation.

\section{Acknowledgement}

The work of B.D. is supported in part by the DOE grant DE-FG02-95ER40917. The work of K.S. is supported by NSF under grant PHY-0505757, PHY05-51164.


\appendix

\section{The Modulus Sector}

As an example of the modulus sector, we consider a KKLT-type \cite{Kachru:2003aw} stabilization scheme in type IIB string theory.

The essential elements in a KKLT-type model are: $(1)$ background fluxes on a type IIB Calabi-Yau three fold giving a Gukov-Vafa-Witten superpotential contribution that fixes complex structure moduli, and $(2)$ gaugino condensation on $D7$ branes or Euclidean $D3$ instantons giving a non-perturbative superpotential contribution that fixes the Kahler moduli. An additional contribution to the scalar potential coming from anti-${D}3$ branes then lifts the solution to a de Sitter vacuum.

The Kahler potential and superpotential in the modulus sector are given by
\begin{equation}
K=-3 \ln (T + \bar T) ~~~,~~~ W= W_{\rm flux} + A e^{-a T} \;.
\end{equation}
Here $T$ is the Kahler modulus, which is related to the compactification radius $R$, $W_{\rm flux}$ is the Gukov-Vafa-Witten piece induced by the fluxes, $A$ is a function of complex structure moduli, the dilaton, and open string fields, and $a$ is related to the beta function of gaugino condensation on the $D7$ branes, $a=2\pi/N_c$ for $SU(N_c)$.

The scalar potential is given by
\begin{equation}
V = e^K \left(K^{I {\bar J}}D_I W D_{{\bar J}}W^*- 3|W|^2\right) + V_{\rm lift}.
\end{equation}
The lifting potential due to the presence of the anti-$D3$ brane is
\begin{equation}
V_{\rm lift}= {D \over (T + \bar T)^{n}} \;,
\end{equation}
with $n$ being an integer ($n=2$ in the original KKLT version) and $D$ is a tuning constant allowing to obtain a Minkowski/de Sitter vacuum. The $F$-term for the Kahler modulus $T$ is given by
\be
F^{T} = - e^{K/2}(D_{{\bar T}} W)K^{T {\bar T}} \,\,.
\ee
After minimizing the scalar potential, one finds \cite{Choi},
\begin{eqnarray} \label{Frelations1}
m_{3/2} &\simeq& {W_{\rm flux} \over (2 ~{\rm Re} T)^{3/2}} \;,\nonumber\\
m_{T} &\simeq& F^{\bar{T}}_{, T} \simeq a ~{\rm Re}T ~ m_{3/2}\;, \nonumber\\
m_{\rm soft} &\simeq& {F_T \over {\rm Re} T} \sim {m_{3/2} \over a ~ {\rm Re} T} \;, \nonumber \\
a ~ {\rm Re} T &\sim& -\ln \left({m_{3/2} \over M_{\rm P}}\right) \simeq 16 \pi^2 \; , \nonumber \\
\end{eqnarray}
where $_,T$ denotes differentiation with respect to $T$.

The complex structure moduli are fixed at the string scale. Supersymmetry breaking is transmitted to the visible sector by a combination of modulus and anomaly mediations. There is a little hierarchy of scales
\be
m_{T} \sim 16 \pi^2 m_{3/2} \sim 16 \pi^2 m_{\rm soft} ,
\ee
which is advantageous from the point of view of the cosmological moduli problem, since the modulus and the gravitino can both be heavy enough to decay before BBN. Typically, we have taken $m_{3/2} \sim 40$ TeV and $m_{T} \sim 1000$ TeV in this work.

\section{Decay channels of the modulus}

The modulus decays through several channels, determined by dimension five operators in the effective $D=4,$ $N=1$ supergravity~\cite{Endo:2006ix,Nakamura:2006uc}. Below, we outline the basic decay modes for the KKLT-type example we considered in this paper.\\
\\
\noindent
{\it (1) Decays into gauge bosons:} The modulus is the real part of the volume modulus, which appears in the gauge kinetic function in the supergravity Lagrangian. The dimension five operator governing the decay into gauge bosons is
\bea
\mathcal{L}_{T gg} &=& ({\rm Re} f_{ij}) ~ \left( -\frac{1}{4}\mathcal{F}_{\mu\nu}^i \mathcal{F}^{\mu\nu j}\right) \nonumber \\
&=& \frac{-1}{4M_{\rm P}} ~ \langle{\rm Re} f_{ij}\rangle ~ \langle{\rm Re} f_{ij, T}\rangle ~ T \mathcal{F}_{\mu\nu}^i \mathcal{F}^{\mu\nu j}
\eea
where $i,~j$ are gauge indices. After canonically normalizing the gauge fields and the modulus, the decay rate for the process $T \rightarrow gg$ is
\be \label{GammaTgg}
\Gamma_{T \rightarrow {\rm gauge}} = \kappa ~ \frac{N_g}{128\pi} ~ \frac{m_{T}^3}{M^2_{\rm P}}
\ee
where $\kappa = \langle{\rm Re} f_{ij}\rangle^{-2} \langle{\rm Re} f_{ij, T}\rangle^2 K_{T\bar{T}}^{-1}$ and $N_g = 12$ is the number of gauge bosons. In the absence of brane magnetic flux, one obtains ${\rm Re} f = {\rm Re} T$, $K_{T\bar{T}} = (3/4){\rm Re}T^2$, and hence $\kappa \sim \mathcal{O}(1)$.\\
\\
\noindent
{\it (2) Decays into gauginos:} The relevant terms in the supergravity Lagrangian are
\bea \label{gaugino}
\mathcal{L}_{T \lambda \lambda} &=& {\rm Re}f_{ij} \left( -\frac{1}{2} \bar{\lambda}^i\slashed{\mathcal{D}} \lambda^j  \right) \nonumber \\
&+& \frac{1}{4}e^{K/2} ~ (D_T W) ~ K^{T\bar{T}} ~ f^{*}_{ij, \bar{T}}\bar{\lambda}^{i}_R\lambda^{j}_R + {\rm h.c.} \, \nonumber \\
\,
\eea
The decay rate through the kinetic term is suppressed as $(m_{\lambda}/m_{T})^2 \sim \mathcal{O}(1/16\pi^2)$ compared to that in Eq.~(\ref{GammaTgg}), where we have used equations of motion, and hence we neglect it. We now use Eq.~(\ref{Frelations1}) and find
\be
\mathcal{L}_{T \lambda \lambda} \supset \frac{1}{4M_{\rm P}} ~ \langle{\rm Re} f_{ij, T} \rangle ~ m_{T} ~ T \bar{\lambda}^{i}_R\lambda^{j}_R ,
\ee
which results in a decay rate $\Gamma_{T \rightarrow {\rm gaugino}}$ comparable to that in Eq.~(\ref{GammaTgg}).\\
\\
\noindent
{\it (3) Decay to MSSM scalars and fermions:} These decays are governed by the non-renormalizable terms in the Kahler potential $K = \lambda_Q T Q Q^{\dagger}/M_{\rm P}$, where the two matter superfields have opposite chirality and the modulus has been normalized.

After using equations of motion, we find a modulus coupling to the kinetic term for the scalars
\be
\mathcal{L}_{T {\tilde Q} {\tilde Q}^{\dagger}} \supset -\left({T \over M_{\rm P}}\right) \tilde{Q}(\partial^2 {\tilde Q}^{\dagger}).
\ee
This channel gives a decay width
\be \label{GammaTQQKin}
\Gamma^{\rm kinetic}_{T \rightarrow {\rm scalar}} \sim \frac{\lambda_Q^2}{8\pi} \left({m_{\tilde Q} \over m_{T}}\right)^4 \times {m_{T}^3 \over M^2_{\rm P}}.
\ee
There are also decays to scalars coming from soft terms in the scalar potential ${m_{\rm soft}}^2 = F^{T}F^{T*}/(T+\bar{T})^2$. The modulus coupling is
\be
\mathcal{L}_{T {\tilde Q} {\tilde Q}^{\dagger}} \sim \frac{m_{3/2}^2}{M_{\rm P}} ~ T \tilde{Q} {\tilde Q}^{\dagger}.
\ee
The decay width through this channel is

\be \label{GammaTQQPot}
\Gamma^{{\rm potential}}_{T \rightarrow {\rm scalar}} \sim \frac{\lambda_Q^2}{8\pi} \left({m_{3/2} \over m_{T}}\right)^4 \times {m_{T}^3 \over M^2_{\rm P}} .
\ee
The decay to MSSM fermions receives chiral suppression and goes as
\be \label{GammaTPsiQ}
\Gamma_{T \rightarrow {\rm fermion}} \sim \frac{\lambda_Q^2}{8\pi} \left({m_{\psi} \over m_{T}}\right)^2 \times {m_{T}^3 \over M^2_{\rm P}} \,\, .
\ee
%
\\
\noindent
{\it (4) Decays into color triplets $X,~{\bar X}$:} These decays can proceed through a non-renormalizable operator in the Kahler potential $K = \lambda_X T^{\dagger} X \bar{X}/M_{\rm P} + {\rm h.c.}$, which results in:
\be
\mathcal{L}_{T X \bar{X}} \supset \frac{\lambda_X}{M}\left((\partial^2 T^{\dagger}) \tilde{X} \tilde{\bar{X}} + F^{T*} \psi_X \psi_{\bar{X}} + \ldots + {\rm h.c.}\right) ,
\ee
and, after using the equation of motion for $T$ and Eq.~(\ref{Frelations1}), gives a decay width
\be \label{GammaTXX1}
\Gamma_{T \rightarrow {\rm scalar}} \sim \Gamma_{T \rightarrow {\rm fermion}} \sim \frac{\lambda_X^2}{8\pi} ~ \frac{m_{T}^3}{M^2_{\rm P}}.
\ee
%
There are other decay channels, similar to the MSSM scalars and fermions, which are suppressed.
\\
\\
\noindent
{\it (5) Decay into MSSM singlets $N$:} These decays can proceed through the non-renormalizable operator in the Kahler potential $K = \lambda_N T^{\dagger} N N/M_{\rm P} + {\rm h.c.}$. The expression for the width is similar to the case of the fields $X$.\\
\\

\noindent
{\it (6) Decay to gravitinos:} The modulus-gravitino interaction terms are obtained by expanding the gravitino bilinear terms in the supergravity Lagrangian in powers of the Kahler modulus $T$ (after making a field-dependent chiral transformation of the gravitinos). The final terms are
\begin{eqnarray}
\mathcal{L} &=& \frac{1}{4} \epsilon^{k\ell mn} \left( G_{,T} \partial_k T - G_{,T^*} \partial_k T^{*} \right)
\bar\psi_\ell \bar\sigma_m \psi_n \nonumber \\
&-& \frac{1}{2} e^{G/2} \left( G_{,T} T + G_{,T^*} T^{*} \right) \left[ \psi_m \sigma^{mn} \psi_n + \bar\psi_m \bar\sigma^{mn} \bar\psi_n \right], \nonumber\\
\end{eqnarray}
where $G =  K +\log |W|^2$ is the kahler function. The decay width to helicity $\pm 1/2$ components is given by
\be \label{GammaTGravitino}
\Gamma_{T \rightarrow {\rm gravitino}} \sim \frac{1}{288\pi} ~ \frac{m_T^3}{M^2_{\rm P}}
\ee
where we have used Eq.~(\ref{Frelations1}), and the modulus has been normalized. The decay width to helicity $\pm 3/2$ components is suppressed by powers of $m_{3/2}/m_{T}$.\\
\\
\\
Summing over all channels, the total decay width of the modulus is found to be
\bea
\Gamma_T \sim \frac{c}{2\pi} ~ \frac{m_T^3}{M^2_{\rm P}}  ~ ~ ~ ~ ~ ~ ~ ~ (c \sim & 0.4) \, ,
\eea
with the branching ratios
\bea
{\rm Br}_{\rm gauge/gaugino} &\sim & 0.98\nonumber \\
{\rm Br}_{\rm scalar/fermion} &\sim & 0.01 \nonumber \\
{\rm Br}_{\rm gravitino} &\sim & 0.01
\eea
Since the main products are gauge bosons and gauginos, we take the final branching ratio to color triplets $X_{\alpha}$ to be ${\rm Br}_{\alpha} \sim 0.1$, as discussed in Section III.




\end{document}